\documentclass{article}
\usepackage{graphicx}

\input{tcilatex}

\begin{document}

\author{Salman Khan\thanks{%
sksafi@phys.qau.edu.pk}, M. Ramzan , M.\ K. Khan \\
Department of Physics Quaid-i-Azam University \\
Islamabad 45320, Pakistan}
\title{ Quantum Model of Bertrand Duopoly}
\maketitle

\begin{abstract}
We present the quantum model of Bertrand duopoly and study the entanglement
behavior on the profit functions of the firms. Using the concept of optimal
response of each firm to the price of the opponent, we found only one Nash
equilibirum point for maximally entangled initial state. The very presence
of quantum entanglement in the initial state gives payoffs higher to the
firms than the classical payoffs at the Nash equilibrium. As a result the
dilemma like situation in the classical game is resolved.\newline
\end{abstract}

In economics, oligopoly refers to a market condition in which sellers are so
few that action of each seller has a measurable impact on the price and
other market factors \cite{Gibbons}. If the number of firms competing on a
commodity in the market is just two, the oligopoly is termed as duopoly. The
competitive behavior of firms in oligopoly makes it suitable to be analyzed
by using the techniques of game theory. Cournot and Bertrand models are the
two oldest and famous oligopoly models \cite{cournot, Betrand}. In Cournot
model of oligopoly firms put certain amount of homogeneous product
simultaneously in the market and each firm tries to maximize its payoff by
assuming that the opponent firms will keep their outputs constant. Later on
Stackelberg introduced a modified form of Cournot oligopoly in which the
oligopolistic firms supply their products in the market one after the other
instead of their simultaneous moves. In Stackelberg duopoly the firm that
moves first is called leader and the other firm is the follower \cite%
{Stackelberg}. In Bertrand model the oligopolistic firms compete on price of
the commodity, that is, each firm tries to maximize its payoff by assuming
that the opponent firms will not change the prices of their products. The
output and price are related by the demand curve so the firms choose one of
them to compete on leaving the other free. For a homogeneous product, if
firms choose to compete on price rather than output, the firms reach a state
of Nash equilibrium at which they charge a price equal to marginal cost.
This result is usually termed as Bertrand paradox, because practically it
takes many firms to ensure prices equal to marginal cost. One way to avoid
this situation is to allow the firms to sell differentiated products \cite%
{Gibbons}.

For the last one decade quantum game theorists are attempting to study
classical games in the domain of quantum mechanics [$5$-$14$]. Various
quantum protocols have been introduced in this regard and interesting
results have been obtained [15-25]. The first quantization scheme was
presented by Meyer \cite{Meyer} in which he quantized a simple penny flip
game and showed that a quantum player can always win against a classical
player by utilizing quantum superposition.

In this letter, we extend the classical Bertrand duopoly with differentiated
products to quantum domain by using the quantization scheme proposed by
Marinatto and Weber \cite{Marinetto}. Our results show that the classical
game becomes a subgame of the quantum version. We found that entanglement in
the initial state of the game makes the players better off. Before
presenting the calculation of quantization scheme, we first review the
classical model of the game.

Consider two firms A and B producing their products at a constant marginal
cost $c$ such that $c<a$, where $a$ is a constant. Let $p_{1}$ and $p_{2}$
be the prices chosen by each firm for their products, respectively. The
quantities $q_{A}$ and $q_{B}$ that\ each firm sells is given by the
following key assumption of the classical Bertrand duopoly model%
\[
q_{A}=a-p_{1}+bp_{2}
\]%
\begin{equation}
q_{B}=a-p_{2}+bp_{1}  \label{E1}
\end{equation}%
where the parameter $0<b<1$ shows the amount of one firm's product
substituted for the other firm's product. It can be seen from Eq. (\ref{E1})
that more quantity of the product is sold by the firm which has low price
compare to the price chosen by his opponent. The profit function of the two
firms are given by%
\[
u_{A}\left( p_{1},p_{2},b\right) =q_{A}\left( p_{1}-c\right) =\left(
a-p_{1}+bp_{2}\right) \left( p_{1}-c\right)
\]%
\begin{equation}
u_{B}\left( p_{1},p_{2},b\right) =q_{B}\left( p_{2}-c\right) =\left(
a-p_{2}+bp_{1}\right) \left( p_{2}-c\right)  \label{E2}
\end{equation}%
In Bertrand duopoly the firms are allowed to change the quantity of their
product to be put in the market and compete only in price. A firm changes
the price of its product by assuming that the opponent will keep its price
constant. Suppose that firm B has chosen $p_{2}$ as the price of his
product, the optimal response of firm A to $p_{2}$ is obtained by maximizing
its profit function with respect to its own product's price, that is, $%
\partial u_{A}/\partial p_{1}=0$, this leads to
\begin{equation}
p_{1}=\frac{1}{2}\left( bp_{2}+a+c\right)  \label{E3}
\end{equation}%
Firm B response to a fixed price $p_{1}$ of firm A is obtained in a similar
way and is given by%
\begin{equation}
p_{2}=\frac{1}{2}\left( bp_{1}+a+c\right)  \label{E4}
\end{equation}%
Solution of Eqs.(\ref{E3} and \ref{E4}) lead to the following optimal price
level that defines the Nash equilibrium of the game%
\begin{equation}
p_{1}^{\ast }=p_{2}^{\ast }=\frac{a+c}{2-b}  \label{E5}
\end{equation}%
The profit functions of the firms at the Nash equilibrium become%
\begin{equation}
u_{A}^{\ast }=u_{B}^{\ast }=\left[ \frac{a+c}{2-b}-c\right] ^{2}  \label{E6}
\end{equation}%
From Eq. (\ref{E6}), we see that both firms can be made better off if they
choose higher prices, that is, the Nash equilibrium is Pareto inefficient.

To quantize the game, we consider that the game space of each firm is a two
dimensional Hilbert space of basis vector $|0\rangle $ and $|1\rangle $,
that is, the game consists of two qubits, one for each firm.\ The composite
Hilbert space $\mathcal{H}$\ of the game is a four dimensional space which
is formed as a tensor product of the individual Hilbert spaces of the firms,
that is, $\mathcal{H=H}_{A}\mathcal{\otimes H}_{B}$, where $\mathcal{H}_{A}$%
\ $\ $and $\mathcal{H}_{B}$ are the Hilbert spaces of firms A and B,
respectively. To manipulate their respective qubits each firm can have only
two strategies $I$ and $C$. Where $I$ is the identity operator and and $C$
is the inversion operator also called Pauli spin flip operator. If $x$ and $%
1-x$ stand for the probabilities of $I$ and $C$ that firm A applies and $y$,
$1-y$, are the probabilities that firm B applies, then the final state $\rho
_{f}$ of the game is given by \cite{Marinetto}
\begin{eqnarray}
\rho _{f} &=&xyI_{A}\otimes I_{B}\ \rho _{i}\ I_{A}^{\dag }\otimes
I_{B}^{\dag }+x\left( 1-y\right) I_{A}\otimes C_{B}\ \rho _{i}\ I_{A}^{\dag
}\otimes C_{B}^{\dag }  \nonumber \\
&&+y\left( 1-x\right) C_{A}\otimes I_{B}\ \rho _{i}\ C_{A}^{\dag }\otimes
I_{B}^{\dag }  \nonumber \\
&&+\left( 1-x\right) \left( 1-y\right) C_{A}\otimes C_{B}\ \rho _{i}\
C_{A}^{\dag }\otimes C_{B}^{\dag }  \label{E7}
\end{eqnarray}%
In Eq. (\ref{E7}) $\rho _{i}=|\psi _{i}\rangle \langle \psi _{i}|$ is the
initial density matrix with initial state $|\Psi _{i}\rangle $, which is
given by%
\begin{equation}
|\psi _{i}\rangle =\cos \gamma |00\rangle +\sin \gamma |11\rangle  \label{E8}
\end{equation}%
where $\gamma \in \left[ 0,\pi \right] $ and represents the degree of
entanglement of the initial state. In Eq. (\ref{E8}) the first qubit
corresponds to firm A and the second qubit corresponds to firm B. The moves
(prices) of the firms and the probabilities $x$, $y$ of using the operators
can be related as follows,%
\begin{equation}
x=\frac{1}{1+p_{1}},\qquad \qquad \qquad y=\frac{1}{1+p_{2}}  \label{E9}
\end{equation}%
where the prices $p_{1}$ and $p_{2}\in \lbrack 0,\infty )$ and the
probabilities $x$, $y\in \left[ 0,1\right] $. By using Eqs. (\ref{E7} - \ref%
{E9}), the nonzero elements of the final density matrix are obtained as%
\begin{eqnarray}
\rho _{11} &=&\frac{(\cos ^{2}\gamma +p_{1}p_{2}\sin ^{2}\gamma )}{\left(
1+p_{1}\right) \left( 1+p_{2}\right) }  \nonumber \\
\rho _{14} &=&\rho _{41}=\frac{(1+p_{1}p_{2})\cos \gamma \sin \gamma }{%
\left( 1+p_{1}\right) \left( 1+p_{2}\right) }  \nonumber \\
\rho _{22} &=&\frac{p_{2}\cos ^{2}\gamma +p_{1}\sin ^{2}\gamma }{\left(
1+p_{1}\right) \left( 1+p_{2}\right) }  \nonumber \\
\rho _{23} &=&\rho _{32}=\frac{(p_{1}+p_{2})\cos \gamma \sin \gamma }{\left(
1+p_{1}\right) \left( 1+p_{2}\right) }  \nonumber \\
\rho _{33} &=&\frac{p_{1}\cos ^{2}\gamma +p_{2}\sin ^{2}\gamma }{\left(
1+p_{1}\right) \left( 1+p_{2}\right) }  \nonumber \\
\rho _{44} &=&\frac{p_{1}p_{2}\cos ^{2}\gamma +\sin ^{2}\gamma }{\left(
1+p_{1}\right) \left( 1+p_{2}\right) }  \label{E10}
\end{eqnarray}%
The payoffs of the firms can be found by the following trace operations%
\begin{eqnarray}
u_{A}\left( p_{1},p_{2},b\right) &=&\mathrm{Trace}\left( U_{A}^{\mathrm{oper}%
}\rho _{f}\right)  \nonumber \\
u_{B}\left( p_{1},p_{2},b\right) &=&\mathrm{Trace}\left( U_{B}^{\mathrm{oper}%
}\rho _{f}\right)  \label{E11}
\end{eqnarray}%
where $U_{A}^{\mathrm{oper}}$ and $U_{B}^{\mathrm{oper}}$ are payoffs
operators of the firms, which we define these as%
\begin{eqnarray}
U_{A}^{\mathrm{oper}} &=&\frac{q_{A}}{p_{12}}\left( k_{B}\rho _{11}-\rho
_{22}+\rho _{33}\right)  \nonumber \\
U_{B}^{\mathrm{oper}} &=&\frac{q_{A}}{p_{12}}\left( k_{A}\rho _{11}+\rho
_{22}-\rho _{33}\right)  \label{E12}
\end{eqnarray}%
where $k_{A}=p_{1}-c$, $k_{B}=p_{2}-c$ and $p_{12}=\frac{1}{\left(
1+p_{1}\right) \left( 1+p_{2}\right) }$. By using Eqs. (\ref{E10} - \ref{E12}%
), the payoffs of the firms are obtained as%
\begin{eqnarray}
u_{A}\left( p_{1},p_{2},b\right) &=&(a-p_{1}+bp_{2})[k_{A}\cos ^{2}\gamma
+\{p_{2}+p_{1}(-1-cp_{2}+p_{2}^{2})\}\sin ^{2}\gamma ]  \nonumber \\
u_{B}\left( p_{1},p_{2},b\right) &=&(a-p_{2}+bp_{1})[k_{B}\cos ^{2}\gamma
+\{p_{1}-p_{2}(1+cp_{1}-p_{1}^{2})\}\sin ^{2}\gamma ]  \nonumber \\
&&  \label{E13}
\end{eqnarray}%
One can easily see from Eq. (\ref{E13}) that the classical payoffs can be
reproduced by setting $\gamma =0$ in Eq. (\ref{E13}).

We proceed similar to the classical Bertrand duopoly to find the response of
each firm to the price chosen by the opponent firm. For firm B's price $%
p_{2} $, the optimal response of firm A is obtained by maximizing its own
payoff (Eq. (\ref{E13})) with respect to $p_{1}$. Similarly, the reaction
function of firm B to a known $p_{1}$ is obtained. These reaction functions
can be written as%
\begin{eqnarray}
p_{1} &=&\frac{k_{B}[-1+p_{2}(a+bp_{2})]+[c+p_{2}+2bp_{2}-bp_{2}^{2}k_{B}+a%
\{2-p_{2}k_{B}\}]\cos 2\gamma }{(2-p_{2}k_{B})\cos 2\gamma -2p_{2}k_{B}}
\nonumber \\
p_{2} &=&\frac{k_{A}[-1+p_{1}(a+bp_{1})]+[c+p_{1}+2bp_{1}+bp_{1}^{2}k_{A}+a%
\{2+p_{1}k_{A}\}]\cos 2\gamma }{(2-p_{1}k_{A})\cos 2\gamma +2p_{1}k_{A}}
\nonumber \\
&&  \label{E14}
\end{eqnarray}

The results of Eq. (\ref{E14}) reduce to the classical results given in Eqs.
(\ref{E3} and \ref{E4}) for the initially unentangled state, that leads to
the classical Nash equilibrium. This shows that the classical game is a
subgame of the quantum game.

Now, we discuss the behavior of entanglement in the initial state on the
game dynamics. It can be seen from Eq. (\ref{E14}) that the optimal
responses of the firms to a fixed price of the opponent firm, for a
maximally entangled state, are given by%
\begin{eqnarray}
p_{1} &=&\frac{bp_{2}^{2}+ap_{2}-1}{2p_{2}}  \nonumber \\
p_{2} &=&\frac{bp_{1}^{2}+ap_{1}-1}{2p_{1}}  \label{E15}
\end{eqnarray}%
Solving these equations, we can obtain the optimal price levels and the
corresponding payoffs of each firm. In this case the following four points
are obtained%
\begin{eqnarray}
p_{1}^{\ast }(1) &=&p_{2}^{\ast }(1)=\frac{a+\sqrt{a^{2}+4\beta }}{-2\beta }
\nonumber \\
p_{1}^{\ast }(2) &=&p_{2}^{\ast }(2)=\frac{2}{a+\sqrt{a^{2}+4\beta }}
\nonumber \\
p_{1}^{\ast }(3,4) &=&\frac{2b}{a\sqrt{2+b}\left( \sqrt{2+b}\pm \gamma
\right) }  \nonumber \\
p_{2}^{\ast }(3,4) &=&-\frac{1}{2b}\left[ a\pm \frac{\gamma }{\sqrt{2+b}}%
\right]  \label{E16}
\end{eqnarray}%
where the numbers in the parentheses correspond to the respective points
(the symbols $\pm $ correspond to points $3$ and $4$ respectively). To
verify which point (points) defines the Nash equilibrium of the game, we use
the second partial derivative condition. That is, for Nash equilibrium, the
strategy (point) must be the global maximum of the payoff function, that is,
$\partial ^{2}u_{A(B)}/\partial p_{1(2)}^{2}<0$ and the payoff function at
the point must be higher than the payoff function at the boundary points. It
can easily be verified that this condition is satisfied only at point $1$.
Hence point $1$ defines the Nash equilibrium of the game. The payoffs of the
firms at the Nash equilibrium become%
\begin{eqnarray}
u_{A}(1) &=&u_{B}(1)=\frac{1}{4\beta ^{4}}[a^{4}+2\alpha ^{2}+2a^{2}b\beta
+a^{3}c\beta -a\{\left( \beta -2\right) \beta -3\}c\beta ^{2}  \nonumber \\
&&\qquad +\sqrt{a^{2}+4\beta }(a^{3}+2a\alpha +c\alpha ^{2}+a^{2}c\beta )]
\label{E18}
\end{eqnarray}%
The new parameters introduced in Eqs. (\ref{E16} and \ref{E18}) are defined
as $\beta =b-2$, $\alpha =2-3b+b^{2}$. The payoffs of the firms at the Nash
equilibrium must be real and positive for the entire range of substitution
parameter $b$. This condition for marginal cost $c<1.4$ is satisfied when $%
a\geq 3.5$. The firms' payoffs at the other three points become%
\begin{eqnarray}
u_{A}(2) &=&u_{B}(2)=-\frac{4}{(a+\sqrt{a^{2}+4\beta })^{4}}%
[a^{5}c+a(-1+b)\{(-9+5b)c-2\sqrt{a^{2}+4\beta }\}  \nonumber \\
&&\qquad -a^{3}\{(8-5b)c+\sqrt{a^{2}+4\beta }\}+(-1+b)^{2}(-2+c\sqrt{%
a^{2}+4\beta })  \nonumber \\
&&\qquad +a^{4}(-1+c\sqrt{a^{2}+4\beta })+a^{2}\{6-4c\sqrt{a^{2}+4\beta }%
+b(-4+3c\sqrt{a^{2}+4\beta })\}]  \nonumber \\
u_{A}(3,4) &=&\frac{(1+b)^{2}(a^{2}(2+b)^{3/2}+a(2+b)(b\sqrt{2+b}c\pm \Gamma
)+b(2b\sqrt{2+b}\pm c\Gamma \left( 2+b\right) ))}{(2+b)^{3/2}(a(2+b)\pm
\sqrt{2+b}\Gamma )^{2}}  \nonumber \\
u_{B}(3,4) &=&-\frac{(1+b)^{2}\sqrt{2+b}\left[ 2ac+abc-2b\pm \sqrt{2+b}%
\Gamma c\right] }{4b(2+b)^{5/2}}  \label{E17}
\end{eqnarray}%
where $\Gamma =\sqrt{4b^{2}+a^{2}(2+b)}$.
\begin{figure}[h]
\begin{center}
\begin{tabular}{ccc}
\vspace{-2cm} \includegraphics[scale=1.2]{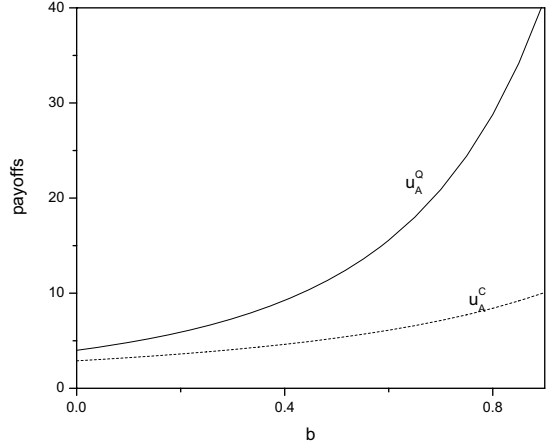}\put(-350,220)(a) &  &
\end{tabular}%
\end{center}
\caption{The payoffs of the firms at the classical and quantum Nash
equilibria against the substitution parameter $b$. The values of the
parameter $a$ and the marginal cost $c$ are chosen as $3.5$ and $0.1$,
respectively. The superscripts $C$ and $Q$ of $u$ represent the classical
and quantum cases, respectively. The subscripts $A$ stands for firm $A$.}
\end{figure}

We present a quantization scheme for the Bertrand duopoly with
differentiated products. To analyze the effect of quantum entanglement on
the game dynamics, we plot the payoffs of the firms at the classical and
quantum Nash equilibria against the substitution parameter $b$ in figure ($1$%
). The values of parameters $a$ and $c$ are chosen to be $3.5$ and $0.1$,
respectively. The solid line ($u_{A}^{Q}(1)$) represents quantum mechanical
payoffs and the dotted line ($u_{A}^{C}$) represents the classical payoffs
of the firms. It is clear from the figure that quantum payoffs of the firms
are higher than the classical payoffs for the entire range of substitution
parameter $b$. The maximum entanglement in the initial state of the game
makes the firms better off. In figure (2), we plot the payoffs of the firms
(Eq. \ref{E17}) against the substitution parameter $b$ at the other three
points which are not the Nash equilibria.
\begin{figure}[h]
\begin{center}
\begin{tabular}{ccc}
\vspace{-2cm} \includegraphics[scale=1.2]{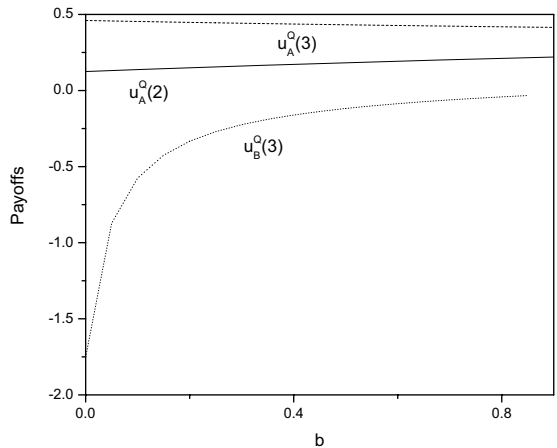}\put(-350,220)(a)
&  &
\end{tabular}%
\end{center}
\caption{The payoffs of the firms at the second and third points as a
function of the substitution parameter $b$. The values of the parameter $a$
and the marginal cost $c$ are chosen as $3.5$ and $0.1$, respectively. The
superscript $Q$ of $u$ represent the classical and quantum cases,
respectively. The subscripts $A$ and $B$ correspond to firms $A$ and $B$
respectively. The numbers in the parentheses represent the corresponding
Nash equilibrium points.}
\end{figure}

In conclusion, we have used the Marinatto and Weber quantization scheme to
find the quantum version of Bertrand duopoly with differentiated products.
We have studied the entanglement behavior on the payoffs of the firms for a
maximally entangled initial state. We found that for large values of
substitution parameter $b,$ both firms can achieve significantly higher
payoffs as compared to the classical payoffs$.$ Furthermore, for maximally
entangled state the quantum payoffs are higher than the classical payoffs
for the entire range of substitution parameter and is the best situation for
both firms. Thus, the dilemma-like situation in the classical Bertrand
duopoly game is resolved.

\begin{center}
{\Huge Acknowledgment}
\end{center}

One of the authors (Salman Khan) is thankful to World Federation of
Scientists for partially supporting this work under the National Scholarship
Program for Pakistan\newline

\end{document}